\documentstyle[epsf,aps,prb]{revtex}

\begin{document}

\twocolumn[\hsize\textwidth\columnwidth\hsize\csname@twocolumnfalse\endcsname

\title{Systematic behaviour of the in-plane penetration
depth in d-wave cuprates}

\author{C.  Panagopoulos, J.R.  Cooper$^*$ and T.  Xiang}

\address{Research Center in Superconductivity, University of
Cambridge, Madingley Road, Cambridge CB3 0HE, United
Kingdom}


\maketitle

\begin{abstract}

We report the temperature T and oxygen concentration
dependences of the penetration depth of grain-aligned ${\rm
YBa_2Cu_3O_{7-\delta} }$ with $\delta$= 0.0, 0.3 and 0.43.
The values of the in-plane $\lambda_{ab}$(0) and
out-of-plane $\lambda_{c}$(0) penetration depths, the low temperature
linear term in $\lambda_{ab}$(T), and the ratio $\lambda_{c}$(0)
/$\lambda_{ab}$(T) were found to increase with increasing $\delta$.
The systematic changes of the linear term in
$\lambda_{ab}$(T) with $T_c$ found here and in recent work on
${\rm HgBa_2Ca_{n-1} Cu_nO_{2n+2+\delta }}$ (n = 1 and 3) $^1$ are discussed.

\end{abstract}

\pacs{PACS number:  }

]

     In a recent study$^1$ of the c-axis coupling of d-wave
high-$T_c$ cuprates we reported the values and temperature (T)
dependences of the in-plane ($\lambda_{ab}$) and
out-of-plane ($\lambda_{c}$) penetration depths for slightly
overdoped$^{2,3}$ ${\rm HgBa_2CuO_{4+\delta} }$ (Hg-1201) with critical temperature
$T_c$ = 93 K and slightly underdoped$^{4,5}$ ${\rm HgBa_2Ca_2Cu_3O_{8+\delta} }$
(Hg-1223) with $T_c$ = 135 K.  For both compounds the low
temperature dependence of $\lambda_{ab}$ was found to be
linear as expected for d-wave superconductivity.  In fact
normalised plots of [$\lambda_{ab}$(0) / $\lambda_{ab}$(T)]$^2$
versus T/$T_c$ were the same, and like ${\rm YBa_2Cu_3O_{7} }$ (YBCO$_7$)$^{6}$
agreed very well with mean field (MF) theory for a weak
coupling d-wave superconductor.  However, recent
angle-resolved photoemission spectroscopy (ARPES)$^{7}$ and
tunneling$^8$ data strongly suggest that for underdoped
samples the superconducting gap, ($\Delta_0$), remains constant, or
even increases slightly, while $T_c$ falls and so large
deviations from MF theory might be expected.  We have
therefore extended our investigation to deoxygenated
(underdoped) pure ${\rm YBa_2Cu_3O_{7-\delta} }$ using the
same ac susceptibility technique to measure the penetration
depth.$^{1,6,9,10}$

     We report experimental results for the values and
temperature dependences of $\lambda_{ab}$ and $\lambda_{c}$ of high
quality c - axis grain-aligned orthorhombic$^{11}$ ${\rm
YBa_2Cu_3O_{7-\delta} }$ (which has two CuO$_2$ planes per unit
cell as well as Cu-O chains) with $\delta$ = 0.0, 0.3 and 0.43, and
compare them with tetragonal$^{12}$ Hg-1201 with one CuO$_2$ plane
per unit cell and tetragonal$^{13}$ Hg-1223 with three CuO$_2$
planes per unit cell.  We find that the presence of the
linear term in $\lambda_{ab}$(T) is independent of the
number of CuO$_2$ planes per unit cell, carrier concentration,
crystal structure, anisotropy and the presence of chains.
Surprisingly our data show good agreement with weak coupling
d- wave theory, and the linear term in [$\lambda_{ab}$(T) /
$\lambda_{ab}$(0)] appears to scale with T /$T_c$.  This result
highlights the need for detailed consideration of the
relationship between superconducting and normal state energy
gaps in underdoped cuprates.

     Sample preparation was carried out by the standard
solid state reaction process using high purity (99.999\%)
Y$_2$O$_3$, BaCO$_3$ and CuO oxides.  Electron probe microanalysis
and x-ray diffraction showed that all samples were single
phase within an accuracy of $\approx$ 1\%.  The fully oxygenated, $\delta$ =
0.0 (YBCO$_7$, $T_c$ = 92 K), samples were prepared by annealing
bulk pieces in pure oxygen atmosphere at 380 $^o$C for 24 hours
and then slowly cooling to room temperature.  (Hereafter $T_c$
represents the temperature where the onset of
superconductivity occurs in the ac susceptibility data for a
measuring field H$_{ac}$ = 3 G rms and frequency f = 333 Hz.)
The $\delta$ = 0.3 (YBCO$_{6.7}$, $T_c$ = 66 K) sample was prepared by
annealing in pure oxygen atmosphere at 650 $^o$C for 12 h and then
quenching in liquid nitrogen, while the $\delta$ = 0.43 (YBCO$_{6.57}$,
$T_c$ = 56 K) sample was annealed in 0.2\% O$_2$ / N$_2$ atmosphere at
550 $^o$C for 12 hours and also quenched into liquid nitrogen.
The final oxygen contents were determined from the weight
change of a fully oxygenated reference sample.  The $\delta$ = 0.0
bulk piece was lightly ground and sedimented in acetone to
obtain a well-defined grain size distribution.  The
sedimented powders were then heat treated to repair any
structural damages to the surface of the grains.$^{14}$ For $\delta$ =
0.3 and 0.43 on the other hand a bulk piece for each $\delta$ was
lightly ground and sieved through a 20 $\mu$m sieve in an argon
glove box to obtain a well-defined grain size distribution$^{10}$
and to avoid surface degradation of the crystallites.$^{14}$ The
collected powders were then kept in argon atmosphere for 30
min before being aligned.  All powders, $\delta$ = 0.0, 0.3 and
0.43, were magnetically aligned in epoxy as described
earlier.$^{1,6,10}$ The average grain diameters corresponding to
the 50\% cumulative volume point were 5 and 10 $\mu$m for the
fully oxygenated and the oxygen deficient samples,
respectively.  The fraction of the unoriented powder in all
grain aligned samples was estimated to be $<$ 5 \%.  Rocking
curve analysis of the $\delta$ = 0.0 and $\delta$ $>$ 0.0 samples gave a
full width at half maximum of $\pm$ 1.4 and 1 $^o$, respectively.$^{15}$ Low
field susceptibility, $\chi$, measurements were performed using
commercial equipment (down to 4.2 K) for samples with $\delta$ =
0.0, 0.3 and 0.43.  The sample with $\delta$ = 0.43 was also
measured down to 1.2 K using a home built susceptometer.
Details of the experimental technique and the application of
London's model for deriving $\lambda$ from the measured $\chi$ in cuprate
superconductors can be found in earlier
publications.$^{1,6,9,10,16}$

     The values of $\lambda_{ab}$(0) derived from our data
are 0.14, 0.21 and 0.28 $\mu$m and the corresponding values for
$\lambda_{c}$(0) are 1.26, 4.53 and 7.17 $\mu$m for $\delta$ = 0.0, 0.3 and 0.43,
respectively.  The errors in $\lambda_{ab}$(0) arising from
a possible uncertainty of $\pm$ 5\% in the alignment can be as high
as $\pm$ 25\%, whereas those in $\lambda_{c}$(0) are $\pm$ 8\%.  However, the
corresponding uncertainty in the linear term in
[$\lambda_{ab}$(T)/$\lambda_{ab}$(0)] is much less, at most
$\pm$ 10\%.  The present results differ from previous work$^{17}$ in
which the surfaces of the particles were probably not as
clean and the degree of grain alignment was probably lower.
As $T_c$ is reduced by lowering the carrier concentration (for
$\delta$ = 0.3 and 0.43), [1 / $\lambda_{ab}$$^2$(0)] falls, a
behaviour which has been extensively discussed in terms of
the Uemura relation.$^{18,19}$ The ratio $\gamma$ = [$\lambda_{c}$(0) /
$\lambda_{ab}$(0)] i.e.  the anisotropy, increases with
oxygen deficiency.

     Figures 1(a) and 1(b), show characteristic low
temperature plots of [$\lambda$(T) / $\lambda$(0)] for the ab- plane
(measured with the applied field H $\parallel$ c) and c-axis
(measured with H $\parallel$ ab), respectively, for the three oxygen
concentrations studied.  The low temperature (T / $T_c$ $<$ 0.25)
linear term in $\lambda_{ab}$(T), is 4.8 $\AA$/K for YBCO$_7$ in
good agreement with that found from microwave measurements
on YBCO$_{6.95}$ single crystals.$^{20}$ As oxygen is removed from the
lattice (the chains) the linear term increases to 12 and 20 $\AA$
/K for $\delta$ = 0.3 and 0.43, respectively.  For YBCO$_7$ we also
observe a linear T dependence in $\lambda_{c}$ at low temperatures but
the relative change is about a factor of two smaller than in
[$\lambda_{ab}$(T) / $\lambda_{ab}$(0)], while $\lambda_{c}$(T) of
YBCO$_{6.7}$ and YBCO$_{6.57}$ obeys a $T^2$ behaviour at low T.  Details
of the systematics of $\lambda_{c}$(T) of cuprate superconductors can
be found in Refs [1,6,21,22].

     In Fig. 2 we present normalised plots of
[$\lambda_{ab}$(0) /$\lambda_{ab}$(T)]$^2$ and [$\lambda_{c}$(0) /$\lambda_{c}$(T)]$^2$
versus T /$T_c$.  There is excellent agreement between the
[$\lambda_{ab}$(0) / $\lambda_{ab}(T)]^2$ curves for the
three oxygen concentrations.  The data in Fig.  2(a) are
compared with the weak coupling theory for a d-wave
superconductor (solid line).$^{23}$ It can be seen that the
d-wave curve fits the data very well.  On the other hand the
[$\lambda_{c}$(0) / $\lambda_{c}(T)]^2$ curves do not fit the d-wave curve and also
differ from each other slightly, because of the effect of
the interplane coupling on $\lambda_{c}$(T).$^{1,6,10,21,22}$ We also find
that [$\lambda_{ab}$(0) / $\lambda_{ab}(T)]^2$ of ${\rm
YBa_2Cu_3O_{7-\delta} }$ agrees with that of Hg-1201 and
Hg-1223.$^{1,10}$ The behaviour of [$\lambda_{ab}$(0) /
$\lambda_{ab}(T)]^2$ is generally similar to that of
${\rm YBa_2(Cu_{1-x}Zn_x)_3O_7}$ (x =0.02 and 0.03),$^6$ except at very low
temperatures where a $T^2$ term developed in the Zn doped
samples due to impurity scattering.

     The full temperature dependences of [$\lambda_{ab}$(0)
/ $\lambda_{ab}(T)]^2$ for ${\rm YBa_2Cu_3O_{7-\delta} }$ ($\delta$
= 0.0, 0.3 and 0.43), ${\rm YBa_2(Cu_{1-x}Zn_x)_3O_7}$ (x =0.02 and 0.03),
Hg-1201 and Hg-1223 are also in agreement with recent data
of ${\rm Bi_2Sr_2CaCu_2O_{8+\delta} }$ (Bi-2212)$^{24}$ and ${\rm Tl_2Ba_2CuO_{6+\delta} }$ (Tl-2201)$^{25}$
single crystals measured by a microwave technique with H $\parallel$
c, but they only agree with another set of microwave data (H
$\parallel$ c) for Bi-2212 single crystals$^{26}$ at T /$T_c$ $<$ 0.3.  At
higher temperatures the data in Ref.  [26] deviate from the
weak-coupling d-wave calculation.  Independent evidence for
the scaling behaviour of [$\lambda_{ab}$(0) /
$\lambda_{ab} (T)]^2$ with T /$T_c$ can also be found in a recent
publication by Bonn et al,$^{27}$ who measured the relative
changes in $\lambda$ with temperature for underdoped, optimally
doped and slightly overdoped untwinned ${\rm
YBa_2Cu_3O_{7-\delta} }$ crystals using a microwave
technique and H $\perp$ c.  However, in Ref.  [27] the changes of
[$\lambda_{a,b}$(0) / $\lambda_{a,b}(T)]^2$ for YBCO$_{6.95}$, at high temperatures, and
YBCO$_{6.6}$, over the whole temperature range, were smaller than
ours and closer, at high temperatures, to the Bi-2212 data
in Ref.  [26].  We do not know the precise origin of this
difference but we believe that for weakly coupled layers,
data taken with H $\parallel$ c give the best measure of the
superfluid density.

     Figure 3 shows plots of {1-[$\lambda_{ab}$(0) /
$\lambda_{ab}(T)]^2$} vs T which is equivalent to {[n$_s$(0) -
n$_s$(T)] / n$_s$(0)}, i.e.  the normalised density of
quasiparticle excitations, where n$_s$(T) is the density of
condensed electrons at a temperature T, for all the samples
studied.  It is clear that the linear term in
{1-[$\lambda_{ab}$(0) / $\lambda_{ab}(T)]^2$} increases as $T_c$
decreases.  If we use the standard BCS result for $\lambda (T)$ of a
d-wave superconductor,$^{28}$ 
\begin{equation}
\left( {\lambda (0) \over \lambda (T)}\right)^2 \approx 
1 - {(2 \ln 2) T \over \Delta_0},
\end{equation}
to fit the experimental data shown in Fig.  3 (at T / $T_c$ $<$
0.25) we find that $\Delta_0$ scales approximately with $T_c$ [Fig.  3
(inset)], giving $\Delta_0$ $\approx$ 2$T_c$, a value close to that expected for
weak- coupling superconductivity.$^{28}$ For comparison we also
include data for the s-wave perovskite ${\rm Ba_{0.6}K_{0.4}BiO_3 }$
(BKBO).$^{10}$ The compounds Bi-2212$^{24,26}$ and Tl-2201$^{25}$ would
also give $\Delta_0$ $\approx$ 2$T_c$ on this plot.  The maximum error in the
linear terms i.e.  the values of $\Delta_0$ in Fig.  3 (inset), is
$\pm$ 20 \%.

     The scaling of $\Delta_0$ with $T_c$, Fig.  3 (inset), is in
agreement with early tunneling spectroscopy data$^{29}$ for
several cuprates as a function of carrier concentration,
ranging from the underdoped to the optimally doped regimes.
It is not consistent however, with more recent tunneling$^8$
and ARPES$^7$ results for underdoped cuprates where $\Delta_0$ was
actually found to increase slightly while $T_c$ falls.  There
seems to be two possible ways of accounting this
discrepancy.  One is that the recent spectroscopic
experiments$^{7,8}$ actually measure the normal state gap.  In
this scenario the effect of the normal state gap would be to
leave small pockets of holes whose superconducting
properties are still described reasonably well by MF theory.
The other is similar to a recent phenomenological
approach.$^{30}$ As shown in Fig.  4, it is probable, that within
experimental error, the unnormalised plots of
[1/$\lambda_{ab}(T)^2$], i.e.  n$_s$(T), versus T at low
temperatures are parallel for samples with different $T_c$
values.  This would correspond to the same number of excited
quasiparticles, [n$_s$(0) - n$_s$(T)], at a given temperature for
all $T_c$ values- as implied by specific heat work on
underdoped YBCO.$^{31}$ Such parallel shifts give [n$_s$(0) - n$_s$(T)]
= aT, where a is independent of doping level ($T_c$).  In
combination with the well-known Uemura relation n$_s$(0) $\propto$ $T_c$,$^{18}$
this gives [1- n$_s$(T) /n$_s$(0)] = b T /$T_c$, where b is
independent of $T_c$.  So at low T, [$\lambda_{ab}$(0) /
$\lambda_{ab}(T)]^2$ versus T/$T_c$ would still scale on to a
single curve even when the MF relation $\Delta_0$ / $T_c$ $\approx$ 2, is
strongly violated.

     In conclusion, we have studied $\lambda_{ab}$(T) and
$\lambda_{c}$(T) of high quality grain-aligned ${\rm
YBa_2Cu_3O_{7-\delta} }$ with $\delta$ = 0.0, 0.3 and 0.43.  The
values of $\lambda_{ab}$(0), $\lambda_{c}$(0) and $\gamma$ were found to
increase with oxygen deficiency.  We find that the existence
of the linear term in $\lambda_{ab}$(T) is independent of
the number of CuO$_2$ planes per unit cell, carrier
concentration, crystal structure, anisotropy and the
presence of chains.  If viewed in isolation, all the
penetration depth data presented here and most of the
microwave measurements for H $\parallel$ c appear to be in excellent
agreement with mean field theory for a weak coupling d-wave
superconductor for which $\Delta_0$ / $T_c$ $\approx$ 2.  However, recent
spectroscopic data are more consistent with a different
approach $^{30}$ in which there is a strong interplay between the
superconducting and normal state gaps.  Clearly the
relationship between these two gaps is of crucial importance
for understanding superconductivity in the cuprates.

     We thank J.W.  Loram and P.A. Lee for enlightening discussions and
B.  Mace for his assistance with the powder preparation.
One of us (C.P) would like to thank Trinity College,
Cambridge for financial support.  This work is supported by
E.P.S.R.C of the United Kingdom.

* On leave from the Institute of Physics, The University of
Zagreb, P.O.  Box 304, Zagreb, Croatia.

\section*{references}

1.  C.  Panagopoulos et al., Phys.  Rev.  Lett.  79, 2320
(1997).

2 G.  B.  Peacock, I.  Gameson, and P.  P.  Edwards (in
preparation).

3 Q.  Xiong et al., Phys.  Rev.  B 50, 10 346 (1994).

4 G.  B.  Peacock, I.  Gameson, and P.  P.  Edwards, Adv.
Mater.  9, 240 (1997).

5 A.  Carrington et al., Physica C 234, 1 (1994); C.K.
Subramaniam, M.  Paranthaman and A.B.  Kaiser, Phys.  Rev.
B 51, 1330 (1995).

6 C.  Panagopoulos, J.R.  Cooper, N.  Athanassopoulou and
J.Chrosch, Phys.  Rev.  B 54, R12 721 (1996).

7 See for example:  J.M.  Harris et al., Phys.  Rev.  B 54,
R15 665 (1996).

8 M.  Oda et al., Physica C (in press).

9 A.  Porch et al., Physica C 214, 350 (1993).

10 C.  Panagopoulos et al., Phys.  Rev.  B 53, R2999 (1996).

11R.  Beyers and T.M.  Shaw in "Solid State Physics - Volume
42" edited by H.  Ehrenreich and D.  Turnbull (Academic
press, inc.  London 1989).

12 J.L.  Wagner et al., Physica C 210, 447 (1993).

13 J.L.  Wagner, B.A.  Hunter, D.G.  Hinks and J.D.
Jorgensen, Phys.  Rev.  B 51, 15 407 (1995).

14 C.  Panagopoulos, W.  Zhou, N.  Athanassopoulou and J.R.
Cooper, Physica C 269, 157 (1996).

15 J.  Chrosch et al., Physica C 265, 233 (1996).

16 D.  Shoenberg, Superconductivity (Cambridge University
Press, Cambridge, 1954), p.164.

17 N.  Athanassopoulou, J.R.  Cooper and J.  Chrosch,
Physica C 235-240, 1835 (1994).

18 Y.J.  Uemura et al., Phys.  Rev.  Lett.  66, 2665 (1991).

19 J.  L.  Tallon et al, Phys.  Rev.  Lett.  74, 1008
(1995).

20 W.  N.  Hardy et al., Phys.  Rev.  Lett.  70, 3999
(1993).

21 T.  Xiang and J.M.  Wheatley, Phys.  Rev.  Lett.  77,
4632 (1996).

22 T.  Xiang and J.M.  Wheatley, Phys.  Rev.  Lett.  76, 134
(1996).

23 A.J.  Schofield (private communication).

24 S.F.  Lee et al., Phys.  Rev.  Lett.  77, 735 (1996).

25 D.  Broun et al., Phys. Rev. B (submitted).

26 T.  Jakobs et al., Phys.  Rev.  Lett.  75, 4516 (1995).

27 D.A.  Bonn et al., Czech.  J.  Phys.  46 S6, 3195 (1996).

28 K.  Maki and H.  Won, J.  Phys.  I France 6, 1 (1996).

29 For reviews see:  J.R.  Kirtley, Int.  J.  of Mod.  Phys.
B 4, 201 (1990); T.  Hasegawa, H.  Ikuta and K.  Kitazawa,
Physical Properties of High Temperature Superconductors,
edited by D.M.  Ginsberg (World Scientific, Singapore,
1992), Volume III, chap.  7, p.525.

30 P.A.  Lee and X-G Wen, Phys.  Rev.  Lett.  78, 4111
(1997).

31 J.W.  Loram et al, Phys.  Rev.  Lett.  71, 1740 (1993).

\section*{Figure Captions}

\begin{figure}
\begin{center}
\leavevmode\epsfxsize=8cm
\epsfbox{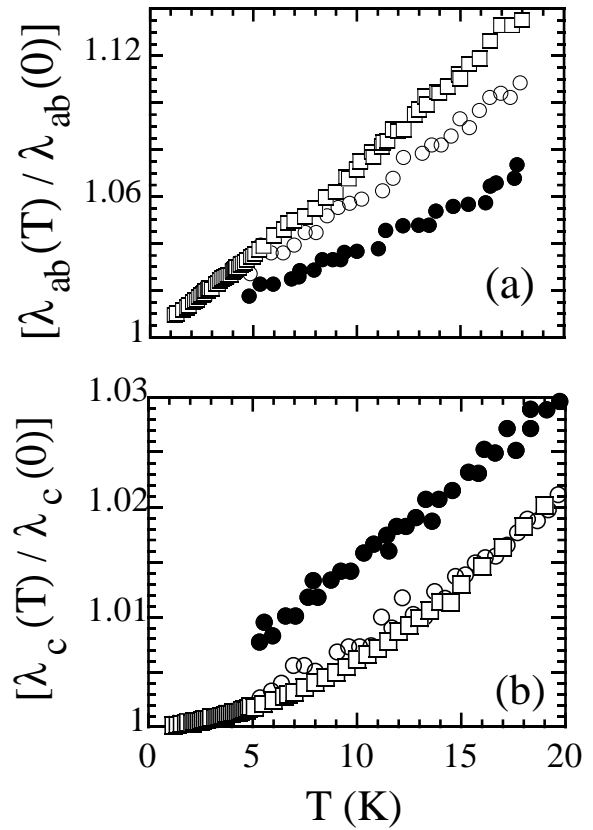}
\caption{Low temperature plots of (a)
[$\lambda_{ab}$(T) / $\lambda_{ab}$(0)] and (b) [$\lambda_{c}$(T) /
$\lambda_{c}$(0)] for YBCO$_7$ (closed circles), YBCO$_{6.7}$ (open circles)
and YBCO$_{6.57}$ (open squares).  The $T_c$, $\lambda_{ab}$(0) and
$\lambda_{c}$(0) values are given in the text.
}
\label{fig1}
\end{center}
\end{figure}

\begin{figure}
\begin{center}
\leavevmode\epsfxsize=8cm
\epsfbox{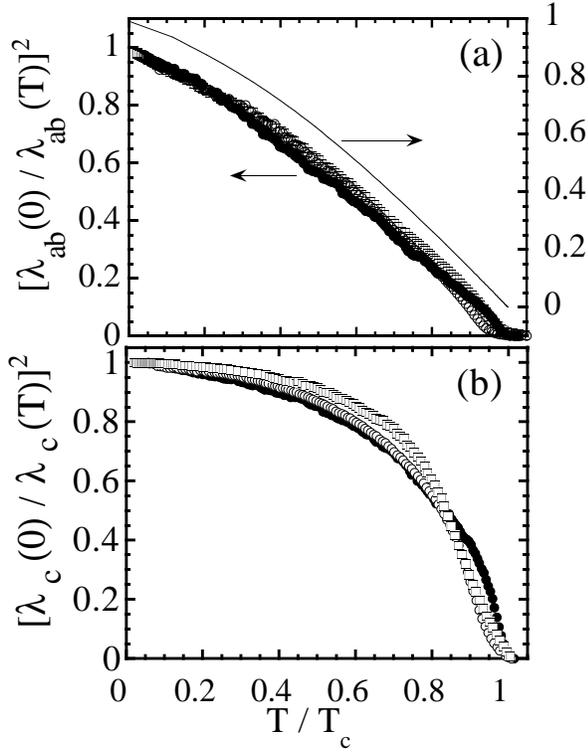}
\caption{Plots of (a) [$\lambda_{ab}$(0) /
$\lambda_{ab}(T)]^2$ and (b) [$\lambda_{c}$(0) / $\lambda_{c}(T)]^2$ as functions of
T /$T_c$ for YBCO$_7$ (closed circles), YBCO$_{6.7}$ (open circles) and
YBCO$_{6.57}$ (open squares).  The solid line in (a) is the
theoretical prediction for the normalised superfluid density
from the weak coupling BCS theory for a d-wave
superconductor.$^{23}$
}
\label{fig2}
\end{center}
\end{figure}

\begin{figure}
\begin{center}
\leavevmode\epsfxsize=8cm
\epsfbox{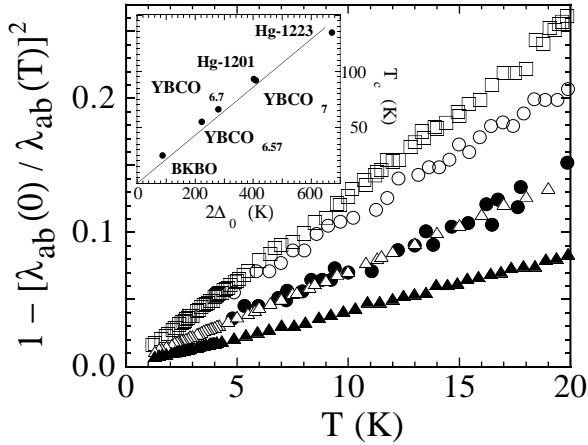}
\caption{Low temperature plot of {1-
[$\lambda_{ab}$(0) / $\lambda_{ab}(T)]^2$} for Hg-1223
[$\lambda_{ab}$(0) = 1770 $\pm$ 300 $\AA$] (closed triangles),$^1$ Hg-1201
[$\lambda_{ab}$(0) 1710 $\pm$ 250 $\AA$] (open triangles),$^1$ YBCO$_7$
(closed circles), YBCO$_{6.7}$ (open circles) and YBCO$_{6.57}$ (open
squares).  Inset:  $T_c$ versus 2$\Delta_0$ as derived from the plot in
the main panel (see text for details).  BKBO is included for
comparison.  The solid line is drawn as a guide to the eye.
}
\label{fig3}
\end{center}
\end{figure}

\begin{figure}
\begin{center}
\leavevmode\epsfxsize=8cm
\epsfbox{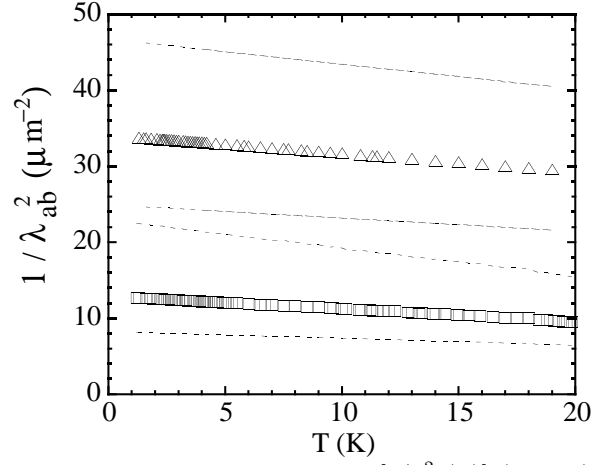}
\caption{Low temperature plot of
[1/$\lambda_{ab}^2$(T)] (i.e.  n$_s$) for YBCO$_{6.57}$ (open
squares) and Hg- 1201 (open triangles) showing the
approximate parallel shift of n$_s$ with $T_c$ as discussed in the
text.  The dashed lines, immediately above and below each
data set, indicate the maximum possible error in
1/$\lambda_{ab}^2$(T) arising from $\pm$ 5\% uncertainty in the
alignment (see text).
}
\label{fig4}
\end{center}
\end{figure}

\end{document}